\documentclass[prl,twocolumn,superscriptaddress,showpacs,amsmath,amssymb]{revtex4-1}
\usepackage{graphicx}
\usepackage{dcolumn}
\usepackage{bm}
\usepackage{color}
\usepackage{epsfig,float,afterpage,amssymb,wrapfig,psfrag}
\usepackage{subfigure}
\newcommand{\bk}{{\bf k}}

\begin{document}

\title{Global Phase Diagram and Momentum Distribution of Single-Particle Excitations in Kondo insulators}
\author{J.~H.~Pixley}
\affiliation{Condensed Matter Theory Center and Joint Quantum Institute, Department of Physics, University of Maryland, College Park, Maryland 20742- 4111 USA}
\author{Rong~Yu}
\affiliation{Department of Physics, Renmin University of China,
Beijing, 100872, China}
\affiliation{Department of Physics and Astronomy, Collaborative Innovation
Center of Advanced Microstructures, Shanghai Jiaotong University,
Shanghai 200240, China}
\author{Silke Paschen}
\affiliation{Institute of Solid State Physics, Vienna University of
 Technology, Wiedner Hauptstra{\ss}e 8-10, 1040 Vienna, Austria}
\author{Qimiao Si}
\affiliation{Department of Physics \& Astronomy, Rice University,
Houston, Texas, 77005, USA}

\date{\today}

\begin{abstract}
Kondo insulators are emerging as a simplified setting to study both magnetic 
and metal-to-insulator quantum phase transitions.
Here, we study a half-filled Kondo lattice model defined on a magnetically frustrated Shastry-Sutherland geometry.
We determine a ``global'' phase diagram that features a variety of zero-temperature phases; these include
Kondo-destroyed antiferromagnetic and paramagnetic metallic phases in addition
to the Kondo-insulator phase.
Our result provides the theoretical basis for understanding how applying 
pressure to a Kondo insulator can close its hybridization gap, liberate the local-moment spins from the 
 conduction electrons, and lead to a magnetically correlated metal.
We also study the momentum distribution of the single-particle excitations in the Kondo insulating state,
 and illustrate how Fermi-surface-like features emerge as a precursor to the actual Fermi surfaces
 of the Kondo-destroyed metals. We discuss the implications of our results for Kondo insulators
 including  SmB$_6$.
\end{abstract}


\maketitle

Quantum criticality in the vicinity of antiferromagnetic order is of interest to a variety of strongly correlated electron systems~\cite{HilbertvH-2010}. Heavy fermion systems occupy a special place in this context~\cite{Si-2010,HeavyFermion-review}.
These systems are typically described by a Kondo lattice model, which contains
 a lattice of local-moment spins coupled to a band of conduction electrons.
 The strong interactions,  which underlie the local moments, allow the ground states of 
 these systems to be readily tuned by external parameters such as pressure or magnetic field.
 For this reason,
antiferromagnetic (AF) quantum critical points (QCPs) have been identified in a host of heavy fermion metals.
Experimental studies at such QCPs ~\cite{Schroder-2000,Paschen-2004,Shishido-2005,Friedemann-2010}
have provided evidence
for 
a Kondo-destruction
local QCP ~\cite{Si-2001,Coleman-2001,Si-2014}, across which the Fermi surface jumps from ``large"
(incorporating the $f$-electrons) to ``small" (excluding the $f$-electrons).
Experimental efforts have also been devoted to
heavy fermion metals that allow a systematic tuning of inherent quantum fluctuations,
through
 strong magnetic frustration ~\cite{Kim-2013,Khalyavin-2013,Mun-2013,Fritsch-2014,Tokiwa-2015}, 
 dimensionality~\cite{Custers-2012} or other related 
 means~\cite{Friedemann-2009,Custers-2010,Jiao-2015},
 which are 
  shedding light on 
 novel phases that involve the development or destruction of not only the AF order parameters
 but also the Kondo entanglement. Such effects have been theoretically considered,
and a global phase diagram has been advanced~\cite{Si-2006,Si2-2010,Coleman-2010,Pixley-2014}.

In the context of such developments in heavy-fermion metals it is then a natural question to ask, whether 
and how quantum phase
transitions can be realized in Kondo insulators~\cite{Si2-2013}? When the local-moment 
spins form a Kondo singlet with the spins of the conduction electrons, 
the resulting Kondo resonances are described in terms of a hybridization 
with the conduction electrons~\cite{Hewson-book}.
An incommensurate electron-filling implies that the chemical potential intersects the hybridized bands,
which leads to a heavy-fermion metallic state.
 By contrast, for a commensurate filling,  the chemical 
potential falls in the middle of a hybridization gap~\cite{Aeppli-1992, Riseborough-2000}, 
and a Kondo-insulator state ensues. 
If, by analogy with the case of the heavy fermion metals, an external parameter 
such as pressure tunes 
the system across a Kondo-destruction transition, the gap of the Kondo insulator will close. 
At the same time, the local-moment 
spins will be liberated from the conduction electrons, 
thereby yielding magnetic states in which the spin-rotational invariance is either spontaneously broken 
({\it e.g.}, an AF order) or preserved (a valence-bond solid or a spin liquid).
This type of qualitative considerations have led to a proposed global phase diagram
for Kondo insulators~\cite{Yamamoto-2010}, but systematic theoretical studies have yet to be performed
 to map this out.

Studies along this direction are also important to  understand
 the on-going experiments on Kondo insulators~\cite{Si2-2013}, 
 which in recent years have been particularly fueled by the search for 
topological 
Kondo insulating states~\cite{Dzero-2010} in SmB$_6$ and related systems.
 Two types of experiments have added to the motivation for the present work.
 First, it is known that applying pressure drives SmB$_6$ from a Kondo insulator to an 
 AF metal~\cite{Barla-2005}, 
 with indications for linear resistivity near the threshold pressure~\cite{Gabani-2003}.  
 Second, torque magnetometry measurements have observed a de Haas-van Alphen
 signal in SmB$_6$~\cite{Li-2014,Tan-2015}, raising the exciting 
 possibility that the Kondo insulator state harbors an incipient Fermi surface
 of the underlying conduction electrons~\cite{Tan-2015}.

In this paper, we study a half-filled Kondo lattice model. First, in order to have a convenient 
tuning parameter for quantum fluctuations of the local-moment magnetism, we focus on the model
defined on the geometrically
frustrated Shastry-Sutherland lattice. 
The calculated phase diagram is strikingly similar to the global phase diagram proposed earlier based on 
qualitative considerations~\cite{Yamamoto-2010}. 
Second, we study the momentum distribution of single-particle excitations
in the Kondo insulating state. 
We identify a Fermi-surface-like feature in the Kondo insulating phase,
and show how it represents a \emph{precursor} to the actual Fermi surfaces
 of the Kondo-destroyed metals that are ``nearby'' in the zero-temperature phase diagram.

The half filled ({\it i.e.} one conduction electron per site)
Kondo-Heisenberg model is defined as
\begin{equation}
H = \sum_{(i,j),\sigma}t_{ij} ( c_{i\sigma}^{\dag}c_{j\sigma}  + \mathrm{h.c.})
+ J_K \sum_i {\bf S}_i\cdot{\bf s}_i^c + \sum_{(i,j)}J_{ij}{\bf S}_i\cdot{\bf S}_j
\label{eqn:ham}
\end{equation}
where $(i,j)$ denotes a sum over neighboring bonds, with a hopping strength $t_{ij}$, 
an RKKY interaction $J_{ij}$, and a Kondo coupling $J_K$.
The spin density of the conduction electrons at site $i$ is
${\bf s}_i^c = c_{i\alpha}^{\dag}(\bm{\sigma}_{\alpha\beta}/2)c_{i\beta}$,
where $\bm{\sigma}_{\alpha\beta}$ are the Pauli spin matrices, 
which are Kondo coupled (via $J_K>0$) to spin-$1/2$ local moments, 
${\bf S}_i=f_{i\alpha}^{\dag} (\bm{\sigma}_{\alpha\beta}/2) f_{i\beta}$. 
To represent the spins, we use fermionic spinons~\cite{Auerbach-book} $f_{i\sigma}$ 
which are subject to the constraint $\sum_{\sigma}f_{i\sigma}^{\dag}f_{i\sigma}=1$.  
The half filled condition is
 $n_c=1$,
where $n_c=\frac{1}{N}\sum_{i,\sigma} \langle c^\dagger_{i\sigma} c_{i\sigma}\rangle$ is the
filling of the conduction band ($N$ is the number of sites in the lattice). 

\emph{Global phase diagram:~}
We consider the two-dimensional (2D) Kondo lattice model on the SSL geometry, 
which is depicted in Fig.~\ref{fig:1}.
Here 
$J_1$ and $J_2$ 
denote the nearest neighbor (NN) and next nearest neighbor (NNN) RKKY interactions,
respectively, and  
$t_1$ and $t_2$ are for the NN and NNN hoppings,  
respectively. In the limit of $J_K=0$ both the Heisenberg model on the SSL~\cite{Shastry-1981,Miyahara-2003} 
and the electronic band structure have been studied~\cite{Shastry-2002,Kariyado-2013} in detail.  
It is known that for $J_2/J_1> 2$ the SSL Heisenberg model exhibits an exact valence bond solid 
(VBS) ground state. In the incommensurate-filling case, quantum phase transitions between the heavy-fermion
metal phase and other magnetic-metal phases have been 
studied~\cite{Pixley-2014,Bernhard-2011,Bernhard-2015}.

We treat the Kondo interaction by introducing a Hubbard-Stratonovich decoupling
 in the 
Kondo-singlet (\emph{i.e.}~hybridization) channel 
$B_i=\sum_{\sigma}c_{i\sigma}^{\dag}f_{i\sigma}$.
In addition, 
to capture the valence bond solid (VBS) and AF orders,
we split the Heisenberg interaction into each of these respective channels 
written as $D_{ij}
= \sum_{\sigma}f_{i\sigma}^{\dag}f_{j\sigma}$ 
and ${\bf S}_i = f_{i\alpha}^{\dag}(\bm{\sigma}_{\alpha\beta}/2)f_{i\beta}$, 
and we weight the channels by a parameter $x$, 
where $0<x<1$~\cite{Pixley-2014}. This leads to the mean field Hamiltonian
\begin{eqnarray}
& &H_{MF} = C
-\sum_{i}(b_i^*c^{\dag}_{i\sigma}
f_{i\sigma} + \mathrm{h.c.})
+\sum_{i}\lambda_if^{\dag}_{i\sigma}f_{i\sigma}
\\
&+&\sum_{(i,j)}\left([(t_{ij} c^{\dag}_{i\sigma}c_{j\sigma}
-Q_{ij}^*f^{\dag}_{i\sigma}f_{j\sigma}) + \mathrm{h.c.}]
+\tilde{J}_{ij}2{\bf M}_i\cdot{\bf S}_j\right)
\nonumber
\label{eqn:Hmf}
\end{eqnarray}
where the sum over $\sigma$ is implied, and we have defined $\tilde{J}_{ij} = (1-x)J_{ij}$.
The constant term is
$C=\sum_i\left(2|b_i|^2/J_K-\lambda_i\right)+
\sum_{(i,j)}(2|Q_{ij}|^2/(xJ_{ij})-\tilde{J}_{ij}{\bf M}_i\cdot{\bf M}_j)$.  The
hybridization is
$b_i=J_K\langle B_i \rangle/2$, and the Hubbard-Stratonovich parameters in the 
resonating valence bond (RVB) 
singlet channel are $ Q_{ij}=xJ_{ij}\langle D_{ij} \rangle/2$.
Lastly, the AF order parameter is $ {\bf M}_i=\langle {\bf S}_i \rangle$ 
[with the ordering wave vector ${\bf Q}=(\pi,\pi)$].
We solve 
for these parameters as described in Ref.~\cite{Pixley-2014} using 
a four site unit cell labelled by $X=A,B,C,D$, with the corresponding four constraints and Kondo hybridization
$\lambda_X$ and $b_X$ respectively. This leads to ten RVB parameters $Q_{x_i},Q_{y_i},Q_{x+ y},$ and $Q_{x-y}$ where $i=1-4$ and the geometry is specified in Ref.~\cite{Pixley-2014}.
We determine the zero temperature phase diagram for various different values of $t_2/t_1$, 
and  fix $x=0.7$. This choice of $x$ is guided by the magnetic phase 
diagram at $J_K=0$, where taking $x=0.7$ captures the quantum AF ground state
 (in the limit of $J_2=0$)~\cite{Pixley-2014}; taking this solution as a candidate state 
 we have checked that our results discussed below
  are robust within a certain range of $x$.

\begin{figure}[t!]
\includegraphics[width=0.75\linewidth]{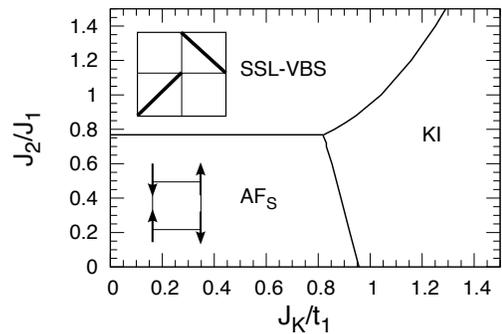}
\caption{Zero temperature phase diagram as a function of frustration ($J_2/J_1$)
and Kondo coupling ($J_K/t_1$), for half filling $n_c=1.0$ and for $t_2/t_1=0.25$. 
The phases and their transitions are described in the main text.
 }
\label{fig:1}
\end{figure}

In Fig.~\ref{fig:1}, we show the phase diagram for $t_2/t_1=0.25$, 
which reveals the three relevant phases.  
For small $J_K/t_1$ and $J_2/J_1$ the model is in the AF phase, 
defined by $0<|{\bf M}| < 1/2$. Here, $Q_{ij}$ is only non-zero along the vertical and 
horizontal bonds and $b_X=0$.
 As this phase has no Kondo screening and is antiferromagnetic we dub it AF$_S$, 
where the $S$ denotes a small Fermi surface.
  In the limit of large frustration and small $J_K/t_1$ the model gives rise to the expected SSL-VBS, where 
  $Q_{x+y}$ and $Q_{x-y}$ are the only non-zero singlet parameters.
   In the limit of large Kondo coupling we find a Kondo insulating (KI) phase. Here
all the $Q_{ij}$ 
   and $b_X$ are non-zero, and they preserve the symmetry of the Shastry-Sutherland lattice.   
  
  We  find direct transitions between AF$_S$ and KI, as well as between SSL-VBS and KI.
 In the incommensurate-filling case~\cite{Pixley-2014}, we found various intermediate 
 phases (between the VBS and heavy Fermi liquid)
  that break the underlying symmetry of the Shastry-Sutherland lattice and 
  as a result exhibit partial Kondo screening. 
  Here, we find these solutions  to be energetically not competitive.
   While our approach yields first order transitions
 (see the Supplementary Material~\cite{supp}),  it is important to see how 
 the fluctuation effects beyond our approach affect the nature of the transitions.
 Based on the studies~\cite{Nica-2013} on the quantum fluctuations in pertinent quantum impurity models 
 in the context of extended dynamical mean field approach~\cite{Si-2001,Si-2014},
 we expect the transitions to be continuous.
 
Two remarks are in order. First, 
the phase diagram 
we have found is strikingly similar to the global phase diagram
proposed for Kondo insulators on qualitative grounds~\cite{Yamamoto-2010}.
From a more general point of view, we can think of $J_2/J_1$ as a measure of the strength of 
quantum fluctuations in the system. Therefore, the phase diagram we have derived
 is representative of Kondo insulating systems more generally; for instance, 
dimensionality tuning could serve for a similar purpose~\cite{Custers-2012}.
Second, each type of phase transition between the KI phase and either the AF$_S$ or the SSL-VBS 
 is actually a \emph{metal to insulator} transition.
Thus, there should be significant effects on Fermi surface probes 
as the transition is approached
 from the metallic side as well as from the insulating side, on which we now turn to.

\begin{figure}[t!]
\centering
\begin{minipage}{.4\textwidth}
\includegraphics[width=0.8\linewidth]{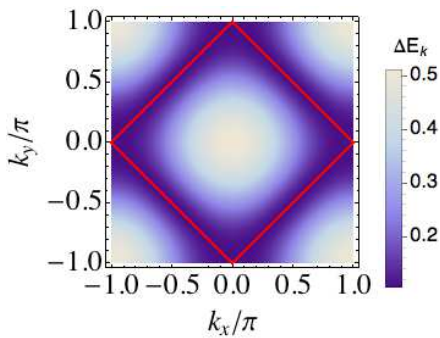}
\end{minipage}
\caption{(Color online). Effects of the underlying Fermi surface on the band gap in the KI phase.
The band gap $\Delta E_{\bk}$ is plotted in momentum space for $J_K/t = 1.5$ in the KI phase.
The minimum of the band gap is along the underlying small Fermi surface (the same as 
for $J_K=0.0$), corresponding to the diamond marked in red. 
}
\label{fig:2}
\end{figure}

\emph{Single-particle excitations and precursor  of the small Fermi surface in the KI phase:~~}  
We focus on the KI phase itself to
consider the momentum distribution of the single-particle excitations.
We are primarily interested in what effects the underlying Fermi surface
 of the conduction electrons have ``imprinted'' on the properties of the insulating state.  
For definiteness, we will denote such small Fermi momenta by ${\bf k}_F$, 
and the large Fermi momenta (those that incorporate the $f$-electrons, see below) by ${\bf k}_F^*$.
 To make concrete 
 connections, we now restrict ourselves  to a model defined on a 2D square lattice with NN couplings only, 
 and therefore consider one site per unit cell. In addition, 
 for the KI phase 
 where the magnetic order parameter vanishes, we no longer consider the effects of the 
 Heisenberg term in Eq. (\ref{eqn:ham}) and set $J_1=J_2=0$;
this
 still keeps the salient properties of the momentum distribution in  this phase. 
 Note that in the following we only have one hopping parameter $t$ and we drop 
 the subscript ``$1$'' in the remainder of the paper.

\begin{figure}[t!]
\centering
\begin{minipage}{.25\textwidth}
\includegraphics[width=1.\linewidth]{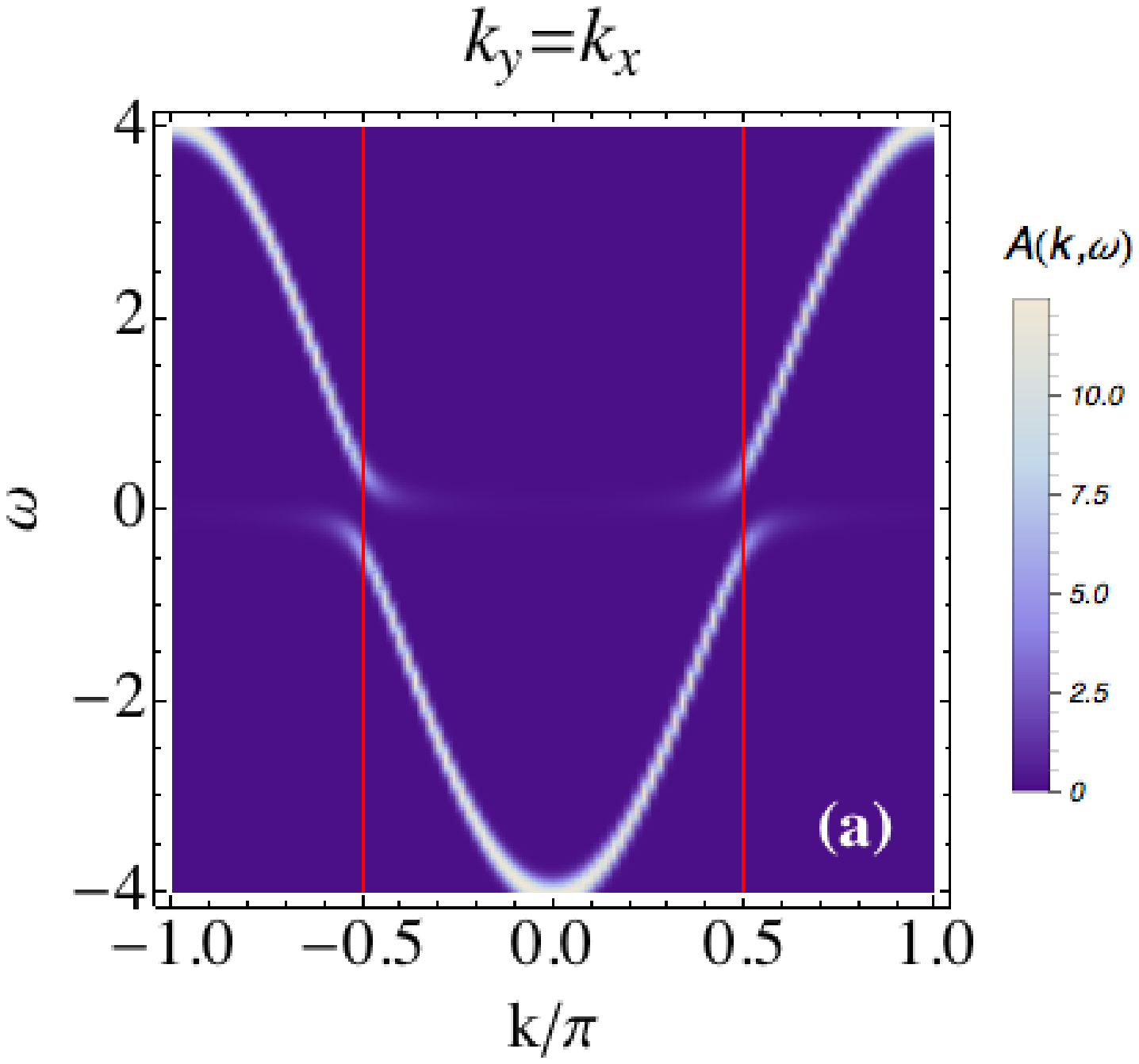}
\end{minipage}
\centering
\begin{minipage}{.25\textwidth}
\includegraphics[width=1.\linewidth]{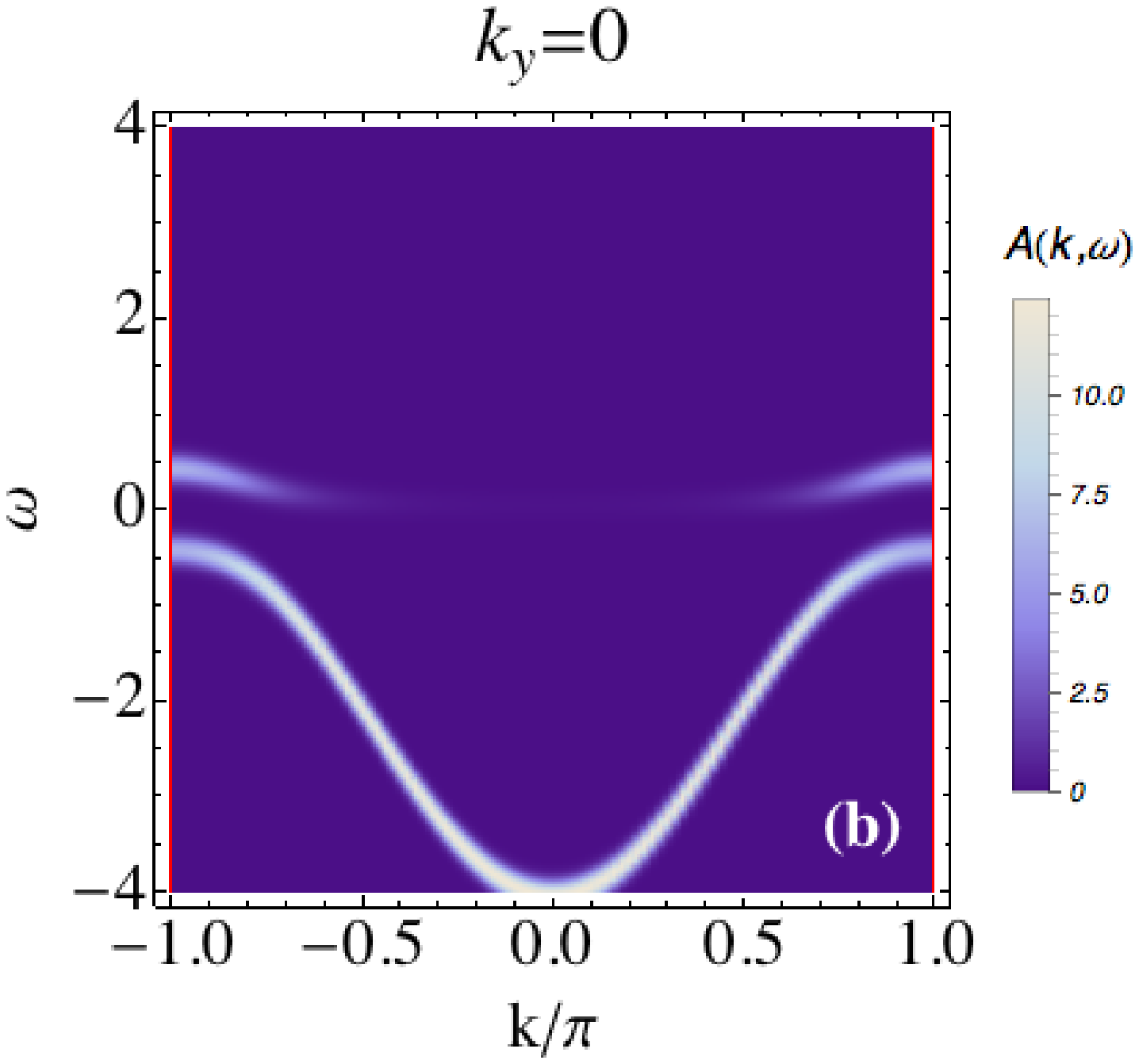}
\end{minipage}
\centering
\begin{minipage}{.25\textwidth}
\includegraphics[width=0.9\linewidth]{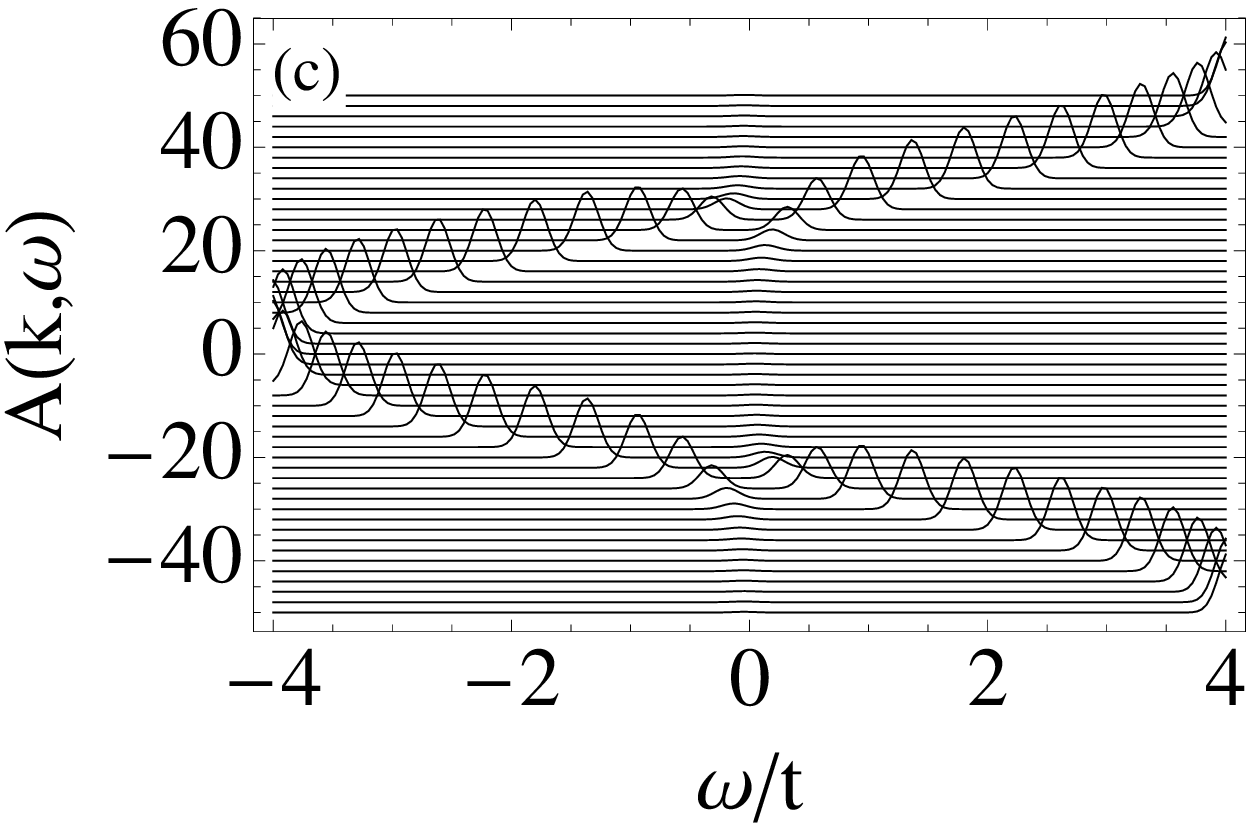}
\end{minipage}
\centering
\begin{minipage}{.25\textwidth}
\includegraphics[width=0.9\linewidth]{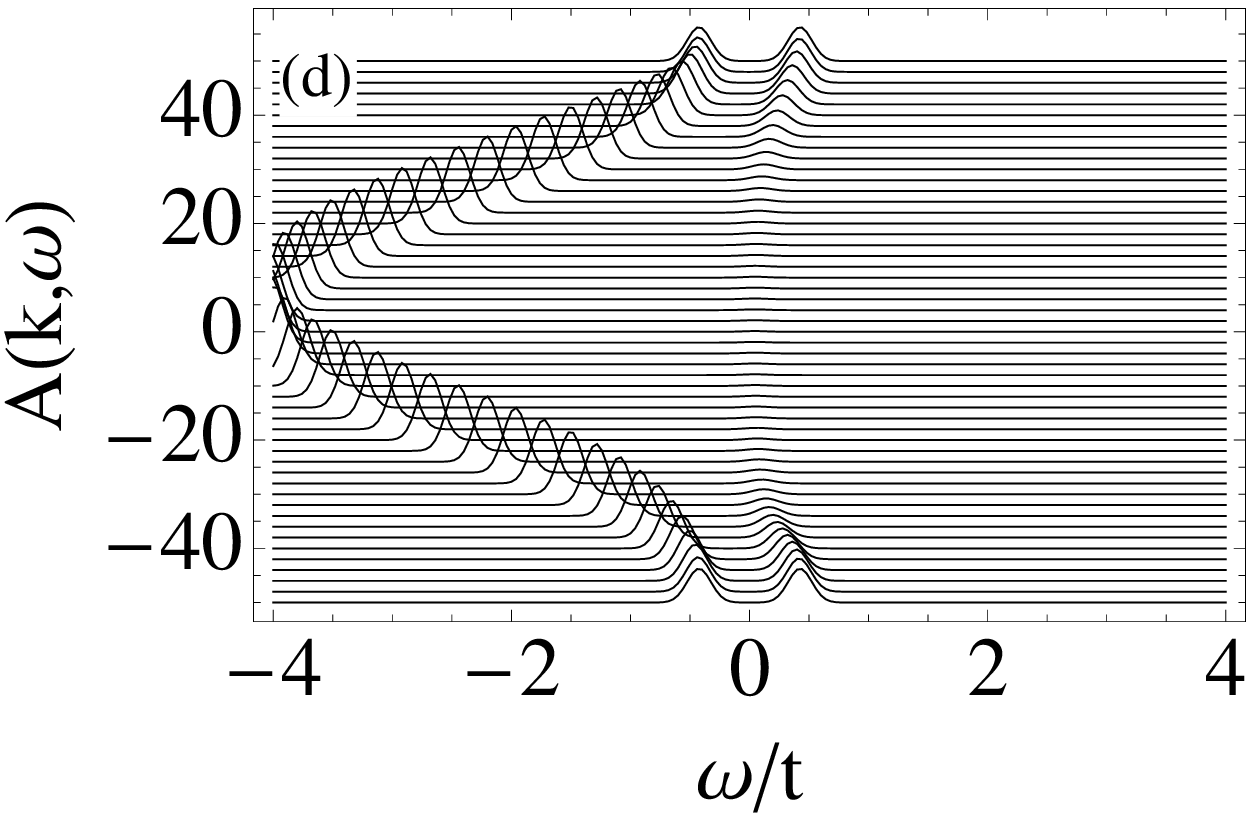}
\end{minipage}
\caption{(Color online).
The spectral function for $J_K/t_1=1.5$ along the content momentum cuts
 (a) $k=k_y=k_x$ and (b) $k=k_x$, $k_y =0$, the red lines mark the location of the small Fermi momentum $k_F$.
  Energy dispersion curves 
of the spectral function are shown along the momentum cuts (c) $k=k_y=k_x$ and (d) $k=k_x$,  $k_y=0$. Here,
 each curve is shifted vertically by an integer corresponding to the wavenumber 
 index and we have broadened the delta function by a Gaussian of width $0.1t$.}
\label{fig:3}
\end{figure}

We first consider the hybridization gap $\Delta E_{{\bf k}}$ as a function of momentum.
Solving for the band structure in the square lattice case yields two bands 
$
E_{{\bf k}\pm} = \frac{1}{2}(\epsilon_{{\bf k}} - \lambda) \pm \sqrt{\left(\frac{\epsilon_{{\bf k}} 
+ \lambda}{2}\right)^2 + b^2}
$,
where $\epsilon_{{\bf k}}=-2t(\cos k_x+\cos k_y)-\mu$ is the dispersion
with a bandwidth $W = 8t$ and chemical potential $\mu$.  
A Kondo insulator of course has no Fermi surface. However,
plotting
the direct hybridization gap 
$\Delta E_{{\bf k}} \equiv E_{{\bf k}+}-E_{{\bf k}-} $
as a function of ${\bf k}$ in the entire Brillouin zone reveals a special surface (line)
in the Brillouin zone. As shown in Fig.~\ref{fig:2},
$\Delta E_{{\bf k}}$ is minimized along the diamond marked in red, which corresponds to the small Fermi 
surface for the half-filled conduction-electron band, $n_c=1$. 
We therefore reach one of our main results, namely $\Delta E_{{\bf k}}$
is {\it minimized on the small Fermi momenta, ${\bf k}_F$}.
The magnitude of  the direct (and indirect) gap is discussed in the Supplementary 
Material (Ref.~~\cite{supp}).

To further illustrate the role played by the small Fermi momenta ${\bf k}_F$, we consider the
dispersion of the single particle excitations and momentum evolution of the 
conduction-electron spectral function  $A({\bf k},\omega) = -\mathrm{Im}G_c({\bf k},\omega)/\pi$
(see the Supplemental Material~\cite{supp}) as shown in Fig.~\ref{fig:3}.  
Along the cut $k_x=k_y$, as shown in Fig.~\ref{fig:3}(a), 
we find the quasiparticle states dispersing towards the small Fermi momentum
${\bf k}_F=(\pi/2,\pi/2)$ as the Fermi energy is approached, although they are eventually gapped out by 
the hybridization when $\bk$ gets too close to ${\bf k}_F$. This point is also illustrated in
the energy dispersion curves, shown in Fig.~~\ref{fig:3}(c). The same trend is also observed 
for the momentum cut $k_y=0$, Figs.~\ref{fig:3} (b) and (d), although here the small Fermi momentum is
${\bf k}_F=(\pi,0)$, which is located on the zone boundary.

To appreciate the above observations,
we note that the small Fermi momenta ${\bf k}_F$ are special because,
for $J_K=0$, the momentum distribution of the $c$ electrons,
$n_{\bf k}= \sum_{\sigma}\langle c_{{\bf k}, \sigma}^{\dag}c_{{\bf k}, \sigma} \rangle$,
has a jump of exactly $1$ across such momenta. 
 A non-zero Kondo coupling will smear this jump~\cite{Hewson-book},
but this smearing occurs gradually.
Indeed, as shown in Fig.~\ref{fig:4} (a), near the small Fermi momentum $\bk_F$,
for various values of the Kondo coupling in the KI phase,
we see a step function for $J_K=0$ develop into an ``S-shape'' pinned 
at 
$\bk_F$ with increasing $J_K$.
(Without a loss of generality, 
we consider the trajectory in the Brillioun zone $0<k_x=k_y<\pi$.)
This smeared jump in the momentum distribution of the occupation number at ${\bf k}_F$
is caused by the same physics that induces the small excitation gap 
at $\bk_F$ illustrated in
Figs.~\ref{fig:2} and \ref{fig:3}.

\begin{figure}[t!]
\centering
\begin{minipage}{.2375\textwidth}
\includegraphics[width=1.\linewidth]{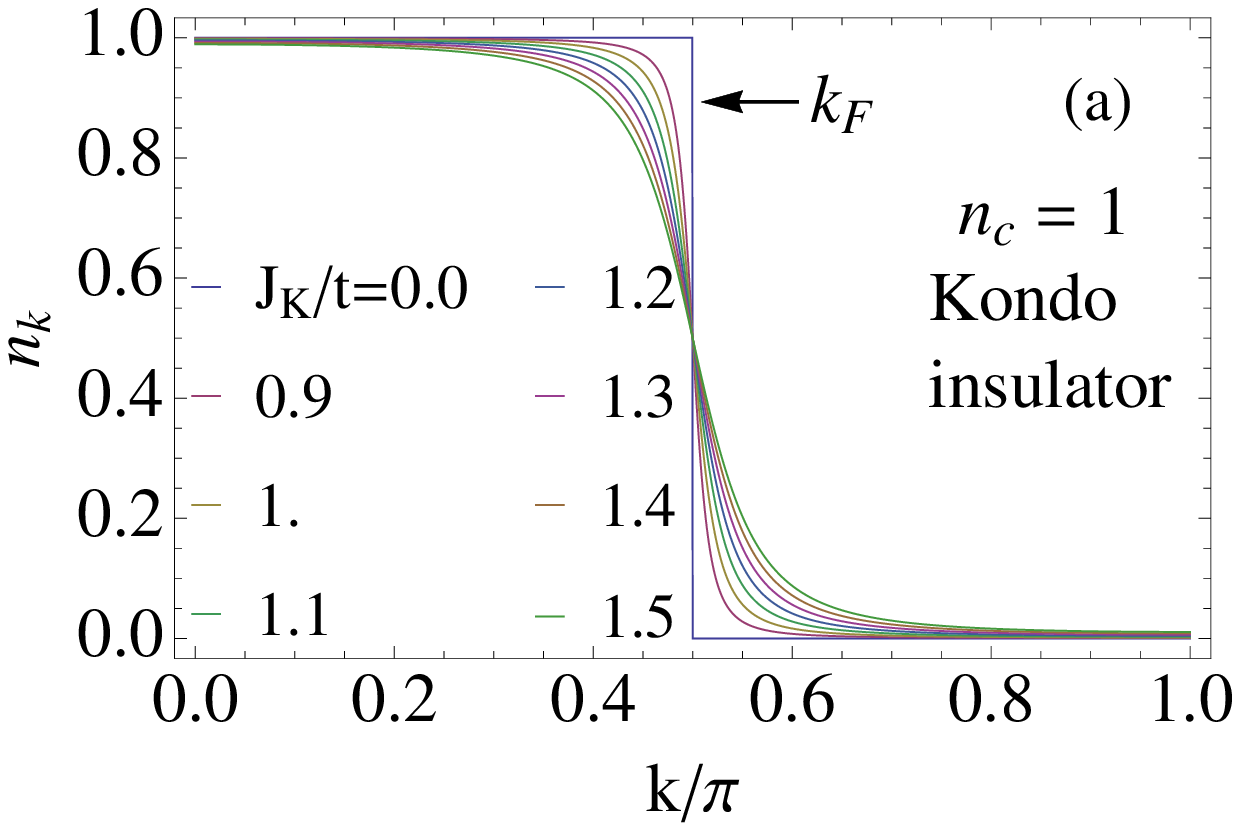}
\end{minipage}
\centering
\begin{minipage}{.2375\textwidth}
\includegraphics[width=1.\linewidth]{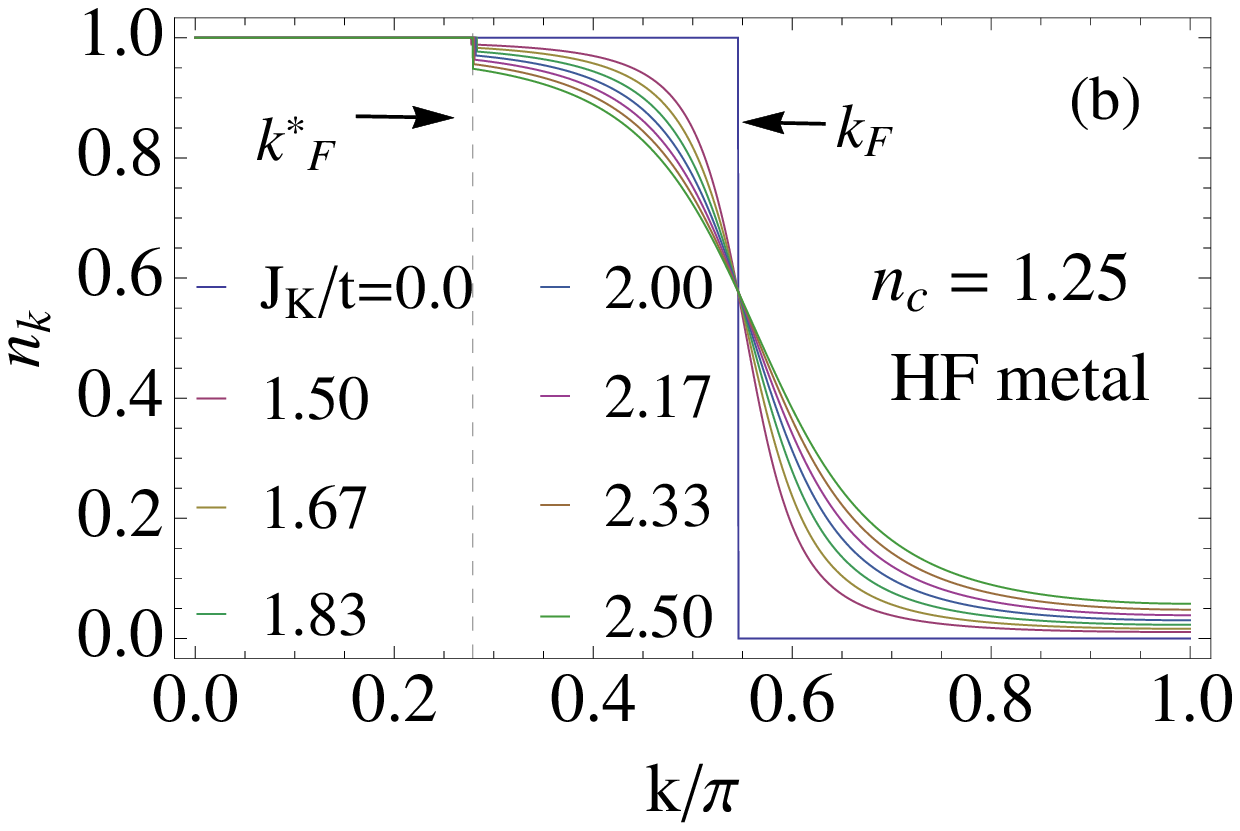}
\end{minipage}
\caption{(Color online).  Momentum distribution of the $c$ electrons $n_k$, for $k=k_x=k_y$.
 $n_k$ versus $k$ for various Kondo couplings in the KI phase (a) and the metallic phase with $n_c=1.25$ (b). 
 The overall shape across $k_F$ in both cases is discussed in the text, 
 while for the metal there is a jump at the large Fermi wave-vector $k_F^*$ (dashed line) of the size of the 
 quasiparticle residue.}
\label{fig:4}
\end{figure}

The simplicity of the Kondo insulator is that the large Fermi momenta, ${\bf k}_F^*$,
are located at the Brillouin zone boundary. This is to be contrasted with the incommensurate-filling case,
as illustrated in Fig.~\ref{fig:4}(b) for $n_c=1.25$. In this case, the features at the small Fermi momenta, 
${\bf k}_F$,
remain similar to the Kondo-insulator case. However, 
now, ${\bf k}_F^*$ occurs in the middle of the Brillouin zone.
As is characteristics of a heavy-fermion metal~\cite{Hewson-book},
$n_{\bf k}$ display a sharp drop at ${\bf k}_F^*$. The drop is tiny,
measuring the exponentially
small quasiparticle residue.

\emph{Discussion and outlook:~~}
We now explore the implications of the global phase diagram shown in Fig.~\ref{fig:1}.
By tuning various Kondo insulating compounds (e.g. under pressure) into and out of various
ground states, this phase diagram opens up the exciting possibility of realizing new types of
 quantum phase transitions.
From the Kondo-insulating state,
such tuning can 
suppress the insulating gap by destroying the Kondo effect,
liberate the local-moment spins, and give rise to
either an
AF or paramagnetic metallic phase. In these metallic phases, the Fermi surface is small as defined earlier,
namely it incorporates only the conduction electrons but not the $f$-electrons.

\begin{figure}[t!]
\includegraphics[height=1.6 in]{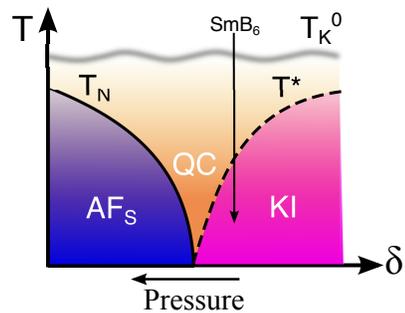}
\caption{(Color online).  Schematic finite temperature phase diagram as a function of the control parameter $\delta$. 
The close proximity of the Kondo-insulator (KI) state to a Kondo-destroyed AF phase (AF$_S$) 
gives rise to a zero temperature QCP and a concomitant  quantum critical (QC) regime.}
\label{fig:0}
\end{figure}

Taking a cut in the global phase diagram
provides the basis to construct a schematic finite temperature phase diagram.
We illustrate this point in Fig.~\ref{fig:0}
on the case when, along the chosen cut,
the
 Kondo-destroyed phase nearby to the Kondo insulator is AF$_S$.
 The AF order is destroyed at the N\'eel temperature $T_N$, 
 whereas the Kondo insulating state persists 
 up to a renormalized effective Kondo energy scale $T^*$, which is distinct from the much higher in energy 
 bare single ion Kondo temperature  $T_K^0$, due to a renormalization 
 from the presence of the RKKY interaction. 
 Thus at finite temperature, by tuning an external control parameter ($\delta$) the system 
 can go from an AF$_S$ 
 phase into a cross over regime driven by the zero temperature quantum critical (QC) fluctuations, 
 which then crosses over to the KI phase at $T^*$.  
Starting from high temperature for a control parameter that
  is in the KI phase at zero temperature, lowering the temperature 
  through $T_K^0$ leads into the QC regime and non-trivial scaling features, 
  which then become 
  cut off by the Kondo-insulator
   state below $T^*$. 
    This provides the theoretical basis to understand
  why pressure induces a Kondo insulator to antiferromagnetic metal 
  transition~\cite{Barla-2005,Gabani-2003,Sun-note}.
 It also makes it at least plausible that a precursor Fermi surface appears in a Kondo insulator,
 similar to an interpretation of the dHvA effect measured in SmB$_6$~\cite{Tan-2015}.
 
 In conclusion, we have determined the global phase diagram in the Kondo-insulator regime 
 of a Kondo-lattice model. This phase diagram makes it natural
 that pressure tuning of Kondo insulators induces magnetic metal phases.
 Our results should also motivate the study of antiferromagnetic and insulator-metal quantum
 phase transitions in Kondo insulators with varying geometrical frustration, or with
 varying dimensionality through thin films or heterostructures.
Finally, we have studied the momentum distribution of the single-electron
 excitations in the Kondo-insulator phase, and demonstrated
 the imprints of a small Fermi surface in this distribution.
 
\emph{Acknowledgements}: We would like to
acknowledge useful discussions with D. T. Adroja, L. Balicas, S. Sebastian, F. Steglich,
A. M. Strydom, L. L. Sun, and H. von L\"ohneysen.
This work was supported in part by JQI-NSF-PFC, LPS-MPO-CMTC, and LPS-CMTC (JHP), 
by  the  National  Science
Foundation of China Grant number 11374361, 
and the Fundamental Research Funds for the Central Universities and the
Research Funds of Renmin University of China (R.Y.),
by the U.S.\ Army Research Office Grant No.\
W911NF-14-1-0497 (S.P.), and by the
U.S. Army Research Office Grant No.\ W911NF-14-1-0525,
 the NSF Grant No.\ DMR-1309531, 
and the Robert A.\ Welch Foundation Grant No.\ C-1411(Q.S.).
J.H.P. acknowledges the hospitality of Rice University. J.H.P., S.P. and Q.S. acknowledge the hospitality 
of the Aspen Center for Physics (NSF Grant no. PHY-1066293) where this work was completed.
The majority of the calculations have been performed on the Shared University Grid at Rice funded by NSF
under Grant EIA-0216467, and a partnership between Rice University, Sun Microsystems,
and Sigma Solutions, Inc..

\bibliography{KI}

\newpage

\onecolumngrid
\setcounter{figure}{0}
\makeatletter
\renewcommand{\thefigure}{S\@arabic\c@figure}
\setcounter{equation}{0} \makeatletter
\renewcommand \theequation{S\@arabic\c@equation}

\title{Global Phase Diagram and Momentum Distribution of Single-Particle Excitations in Kondo insulators}

\author{J.~H.~Pixley}
\author{Rong~Yu}
\author{Silke Paschen}
\author{Qimiao Si}

\maketitle

\section*{{\Large Supplementary Material}}

\begin{figure}[!h]
\centering
\begin{minipage}[b]{20pc}
\includegraphics[height=2.5in,angle=-90]{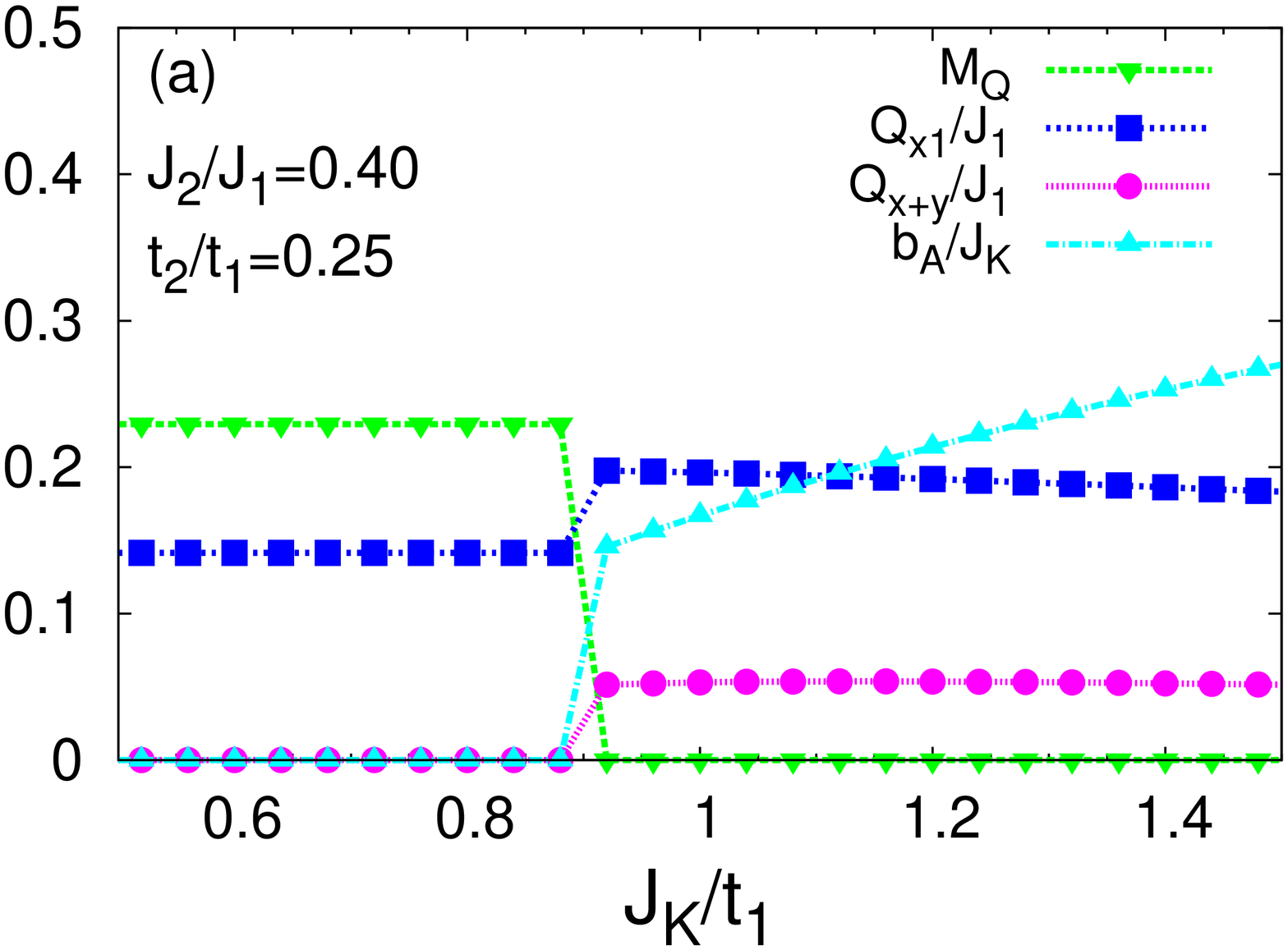}
\end{minipage}
\begin{minipage}[b]{20pc}
\includegraphics[height=2.5in,angle=-90]{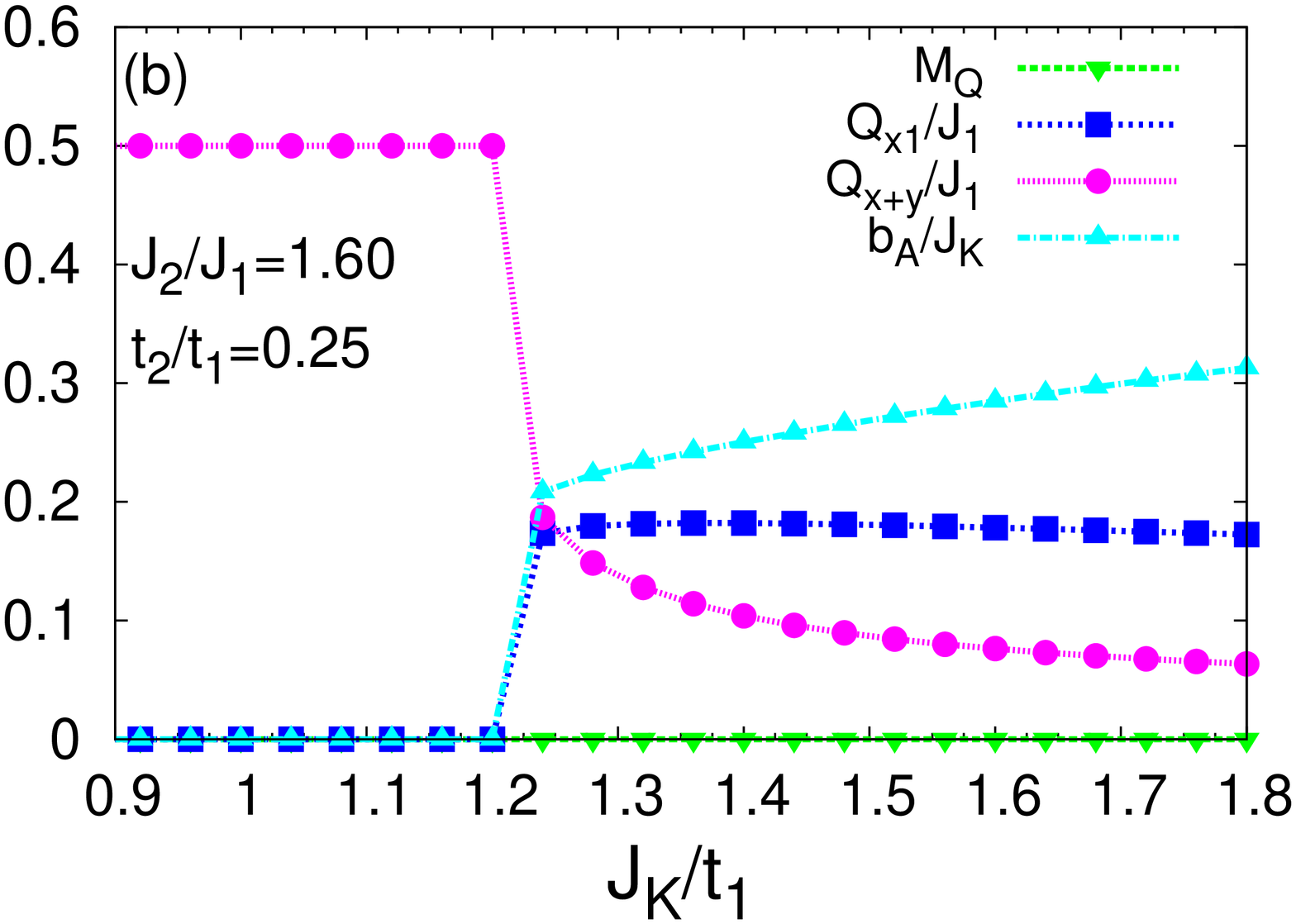}
\end{minipage}
\caption{Mean field parameters across each of the Kondo insulating to Kondo destroyed transitions for $t_2=0.25 t_1$.
(Only the distinct parameters are shown.)
 (a) For the AF$_S$ to KI phase transition as a function of the Kondo coupling $J_K$ for $J_2/J_1=0.4$.  We find a direct first order transition between the AF$_S$ and KI phase signaled by a vanishing of the magnetic order parameter $M_{{\bf Q}}$ (where ${\bf Q}$ corresponds to the AF wave vector);
 (b) Mean field parameters for the SSL-VBS to KI phase transition as a function of the Kondo coupling $J_K$
  for $J_2/J_1=1.6$. We find a direct first order transition between the VBS and KI phase.
  }
\label{fig:S1}
\end{figure}

Here we discuss the various equations we have used in studying the nature of the Kondo insulating phase. 
We show the mean field parameters as a function of the Kondo coupling $J_K$ across each phase transition 
in Fig.~\ref{fig:S1} (as in the phase diagram Fig. 1 of the main text) and also discuss the direct and indirect 
single-particle excitation gaps in the Kondo insulating phase.

\section{Equations for the single particle excitations in the Kondo insulating phase}
Here we give the equations for the occupation number and spectral function within our approach.
In the absence of the RKKY interaction focusing on a square lattice, and after the Hubbard-Stranotovich 
decoupling, the Hamiltonian becomes 
\begin{eqnarray}
H_{MF} &=&\sum_{i,\alpha}i\lambda_i\left(f^{\dag}_{i\alpha}f_{i\alpha} -\frac{1}{2} \right)
\nonumber
+\sum_{\langle i,j\rangle,\sigma}t_{ij}\left(c^{\dag}_{i\sigma}c_{j\sigma} +\mathrm{h.c.}\right)+\frac{2}{J_K}\sum_i |b_i|^2+
\sum_{i,\sigma}\left(b_i^*c^{\dag}_{i\sigma}f_{i\sigma} + \mathrm{h.c.}\right).
\end{eqnarray}
Focusing on translationally invariant solutions, we can set $i\lambda_i=-\lambda$ and $b_i=b$.
We can then diagonalize the Hamiltonian via Fourier transform and obtain
\begin{equation}
H_{MF} = H_C + \sum_{\bk,\sigma} \left[ E_{\bk+}\gamma_{\bk\sigma+}^{\dag}\gamma_{\bk\sigma+} + E_{\bk-}\gamma_{\bk\sigma-}^{\dag}\gamma_{k\sigma-}\right]
\end{equation}
where the light and heavy bands ($+/-$) are
\begin{equation}
E_{\bk\pm} = \frac{1}{2}(\epsilon_{\bk} - \lambda) \pm \sqrt{\left(\frac{\epsilon_{\bk} + \lambda}{2}\right)^2 + b^2},
\end{equation}
and 
\begin{equation}
H_C = N\Big(\frac{2}{J_K}b^2  + \lambda\Big).
\end{equation}
The quasiparticle operators, $\gamma_{\bk \sigma \pm}$, are related to the original degrees of freedom via
\begin{eqnarray}
c_{\bk\sigma}&=& u_{\bk}\gamma_{\bk\sigma+}+v_{\bk}\gamma_{\bk\sigma-},
\\
f_{\bk\sigma} &=& v_{\bk}\gamma_{\bk\sigma+}-u_{\bk}\gamma_{\bk\sigma-},
\end{eqnarray}
where $u_{\bk}^2+v_{\bk}^2=1$
and the form factors are
\begin{eqnarray}
u_{\bk}^2 &=& \frac{1}{2}+\frac{1}{2}\frac{\epsilon_{\bk}+\lambda}{\sqrt{(\epsilon_{\bk}+\lambda)^2+4b^2}},
\\
v_{\bk}^2 &=& \frac{1}{2}-\frac{1}{2}\frac{\epsilon_{\bk}+\lambda}{\sqrt{(\epsilon_{\bk}+\lambda)^2+4b^2}}.
\end{eqnarray}
With these results in hand we can now determine the Green function of the conduction electrons,
\begin{equation}
G_c(\bk,i\omega_n) = \frac{u_{\bk}^2}{i\omega_n-E_{\bk+}}+ \frac{v_{\bk}^2}{i\omega_n-E_{\bk-}}
\end{equation}
with a spectral function
\begin{equation}
A_c(\bk,\omega) = u_{\bk}^2\delta(\omega-E_{\bk+})+ v_{\bk}^2\delta(\omega-E_{\bk-}) .
\end{equation}
Consequently, the occupation number of the $c$ electrons at zero temperature is
\begin{equation}
n_{\bk}  = u_{\bk}^2\Theta(-E_{\bk+})+ v_{\bk}^2\Theta(-E_{\bk-}),
\end{equation}
where $\Theta(x)$ is the Heaviside step function.
From the Green function we can also determine the quasiparticle residue in the heavy Fermi liquid phase for the filling $n_c=1.25$ as in the main text. In the heavy Fermi liquid phase the total filling will be $n_c+n_f=2.25$, hence  $E_{\bk_F^* +}=0$ and there is only a pole at $E_{\bk_F^* -}$ away from zero energy.
Therefore, the residue for the $c$ and $f$ electrons at the Fermi energy is
$z_c = v_{\bk_F^*}^2$,
and 
$z_f = u_{\bk_F^*}^2$ respectively.
\section{Mean Field Parameters}

\section{Direct and Indirect Gaps}

\begin{figure}[!h]
\centering
\begin{minipage}[b]{20pc}
\includegraphics[height=1.6in]{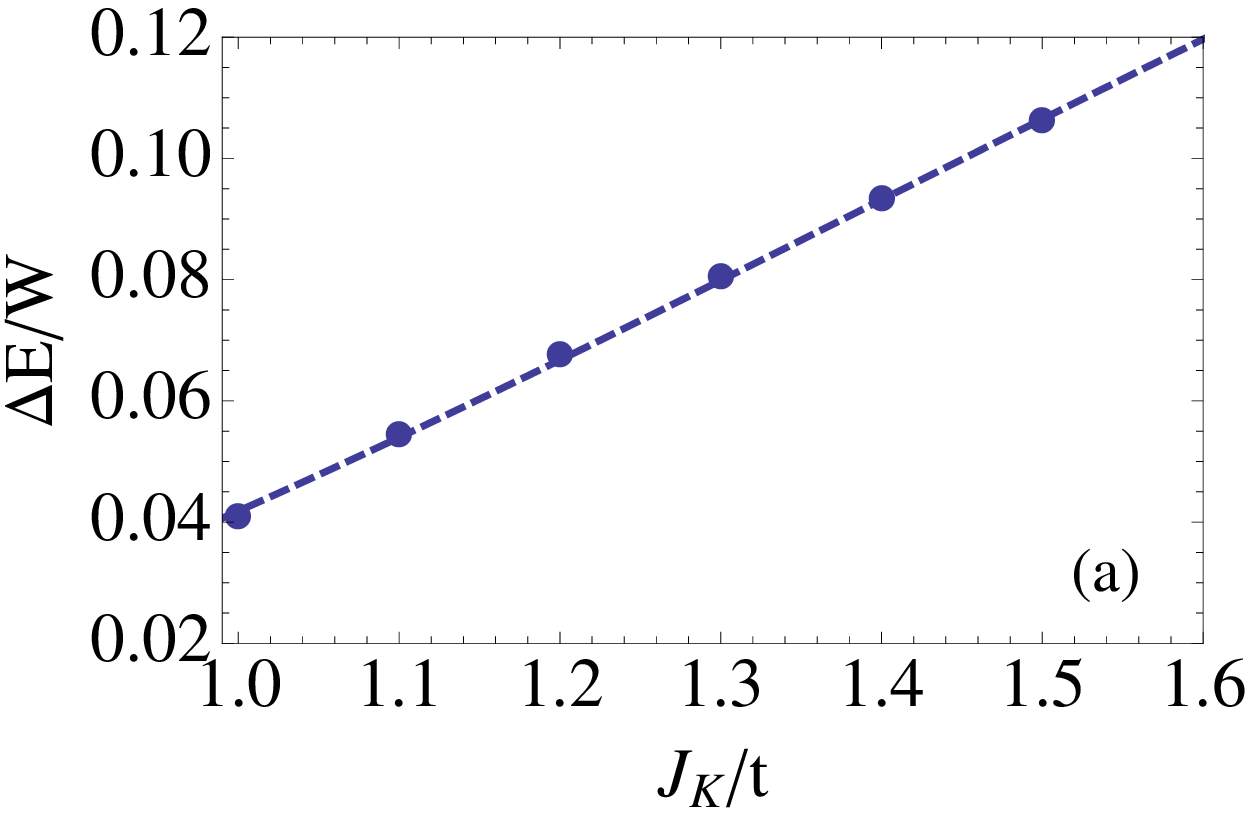}
\end{minipage}
\begin{minipage}[b]{20pc}
\includegraphics[height=1.6in]{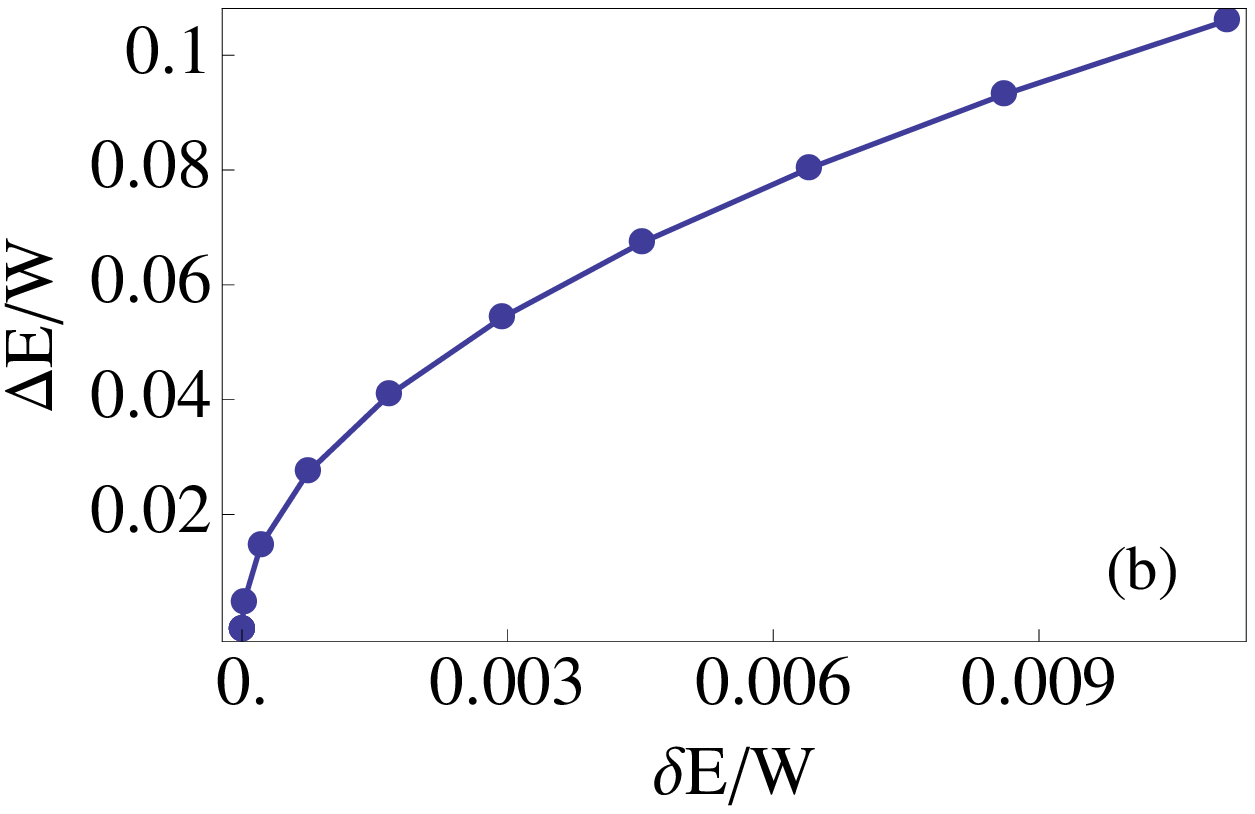}
\end{minipage}
\caption{Direct and indirect gaps in the Kondo insulating phase.  
(a) Direct gap $\Delta E$ as a function
of the Kondo coupling $J_K$. Here, the dashed line is a fit to the expected exponential form [$a \exp(-b/J_K)$]. 
(b) The direct gap as a function of the indirect gap $\delta E$, satisfying $\Delta E \sim \sqrt{\delta E}$.}
\label{fig:S2}
\end{figure}

We define the direct gap
($\Delta E \equiv \mathrm{min}\Delta E_{{\bf k}}  $),
which captures the lowest lying 
excitationsthat don't change momentum.
Likewise, we define the indirect gap 
($\delta E \equiv \mathrm{min}E_{{\bf k}+}-\mathrm{max}E_{{\bf k}-}  $), 
which corresponds to the lowest-energy excitation when the momentum is allowed
to change.
In Fig.~\ref{fig:S2} (a), we show the direct 
gap as a function of the
Kondo coupling $J_K$, which has the expected exponential dependence.
Compared to the indirect gap, which is of the order of the Kondo temperature,
the direct gap is larger and has a square-root dependence on the latter, as is shown
in Fig.~\ref{fig:S2} (b).

\end{document}